\newif\ifmnras
\mnrastrue
\mnrasfalse
\ifmnras
\documentclass[useAMS,usenatbib,usedcolumn]{mnras}
\else
\documentclass[apj,numberedappendix]{emulateapj}
\fi
\usepackage{graphics,epsf}
\usepackage{amsmath}                % American Mathematical Society package
\usepackage{amsfonts}               % American Mathematical Society fonts
\usepackage{amssymb}                % American Mathematical Society symbol
\usepackage{epsfig}                 % EPS figures
\usepackage{graphicx}
\usepackage{float}
\usepackage{amsmath}

\newcommand{\erg}{{~ \mathrm{erg}}}
\newcommand{\zams}{\mathrm{ZAMS}}
\ifmnras
\else
\newcommand{\nar}{{~\rm New Astronomy Reviews}}
\newcommand{\na}{{~\rm New Astronomy}}
\newcommand{\pasa}{{~\rm Publications of the Astronomical Society of Australia}}
\fi

\ifmnras
\title[A He-O shell in CCSN progenitors]{A mixed helium-oxygen shell in some core-collapse supernova progenitors}
\author[R. A. Gofman, A. Gilkis, N. Soker]{Roni Anna Gofman$^{1}$\thanks{Contact e-mail: \href{rongof@campus.technion.ac.il}{rongof@campus.technion.ac.il}}, Avishai Gilkis$^{2}$\thanks{Contact e-mail: \href{agilkis@ast.cam.ac.uk}{agilkis@ast.cam.ac.uk}}, Noam Soker$^{1,3}$\thanks{Contact e-mail: \href{soker@physics.technion.ac.il}{soker@physics.technion.ac.il}}
\\
$^{1}$ Department of Physics, Technion -- Israel Institute of Technology, Haifa 3200003, Israel \\
$^{2}$ Institute of Astronomy, University of Cambridge, Madingley Road, Cambridge, CB3 0HA, UK  \\
$^{3}$ Guangdong Technion Israel Institute of Technology, Shantou, Guangdong Province, China
}
\fi

\begin{document}

\ifmnras
\label{firstpage}
\pagerange{\pageref{firstpage}--\pageref{lastpage}} \pubyear{2018}
\maketitle
\else
\title{A mixed helium-oxygen shell in some core-collapse supernova progenitors}
\author{Roni Anna Gofman\altaffilmark{1}, Avishai Gilkis\altaffilmark{2} \& Noam Soker\altaffilmark{1,3}}
\altaffiltext{1}{Department of Physics, Technion -- Israel Institute of Technology, Haifa
3200003, Israel; rongof@campus.technion.ac.il; }
\altaffiltext{2}{Institute of Astronomy, University of Cambridge, Madingley Road, Cambridge, CB3 0HA, UK; agilkis@ast.cam.ac.uk}
\altaffiltext{3}{Guangdong Technion Israel Institute of Technology, Shantou, Guangdong Province, China}
\fi

\begin{abstract}
We evolve models of rotating massive stars up to the stage of iron core collapse using the \textsc{mesa} code and find a shell with a mixed composition of primarily helium and oxygen in some cases. In the parameter space of initial masses of $13$-$40 M_\odot$ and initial rotation velocities of $0$-$450 ~ \mathrm{km} ~ \mathrm{s}^{-1}$ that we investigate, we find a mixed helium-oxygen (He-O) shell with a significant total He-O mass and with a helium to oxygen mass ratio in the range of $0.5$-$2$ only for a small fraction of the models. While the shell formation due to mixing is instigated by rotation, the pre-collapse rotation rate is not very high.The fraction of models with a shell of He-O composition required for an energetic collapse-induced thermonuclear explosion is small, as is the fraction of models with high specific angular momentum, which can aid the thermonuclear explosion by retarding the collapse. Our results suggest that the collapse-induced thermonuclear explosion mechanism that was revisited recently can account for at most a small fraction of core-collapse supernovae. The presence of such a mixed He-O shell still might have some implications for core-collapse supernovae, such as some nucleosynthesis processes when jets are present, or might result in peculiar sub-luminous core-collapse supernovae.
\ifmnras
\else
\smallskip \\
\textit{Key words:} stars: massive --- stars: rotation --- supernovae: general
\fi
\end{abstract}

\ifmnras
\begin{keywords}
stars: massive --- stars: rotation --- supernovae: general
\end{keywords}
\fi

% ==========================================================
\section{INTRODUCTION}
\label{sec:intro}
% ==========================================================

At the end of their lives massive stars release more than $10^{53} \erg$ of gravitational energy when they explode as core-collapse supernovae (CCSNe). Most of this energy is carried by neutrinos emitted in the transformation of the core into a neutron star (NS). The kinetic energy of the ejecta is only about $10^{50}$ to $\mathrm{few}\times 10^{52} \erg$.

There are two CCSN explosion mechanisms that account for the kinetic energy by utilizing a small fraction of the gravitational energy. The older one is the delayed neutrino heating mechanism (e.g., \citealt{Wilson1985, BetheWilson1985};  \citealt{Janka2012} and \citealt{Mulller2016} for reviews), in which it is assumed that the explosion energy is attained from absorption of a fraction of the neutrino energy in the material surrounding the collapsing core. The second explosion mechanism that utilizes the gravitational energy is the negative jet feedback mechanism (JFM; for a review of the JFM see \citealt{Soker2016Rev}), where the kinetic energy is the result of a part of the material accreting onto the newly formed compact object being ejected in a bipolar outflow termed `jets'. A third mechanism for CCSNe is the collapse-induced thermonuclear explosion (CITE), in which the collapsing material is compressed and heated, amplifying the nuclear reactions up to a thermonuclear runaway \citep{Burbidgeetal1957,KushnirKatz2015}. In this last scenario, the process is initiated by gravitational collapse, but the energy release is from nuclear fusion.

As the mechanism of CCSNe is an open issue and there is no consensus on its nature, all scenarios must be considered and studied. The most well studied is the delayed neutrino heating mechanism, though it cannot account for CCSNe with kinetic energies of $E_\mathrm{SN} \ga 2 \times 10^{51} \erg$ (e.g., \citealt{Fryer2006, Fryeretal2012, Sukhboldetal2016, SukhboldWoosley2016}), and even for weaker explosions it encounters severe problems (e.g., \citealt{Papishetal2015a, Kushnir2015b}). The formidable complication presented by unusually luminous and energetic CCSNe \citep{GalYam2012} has prompted the notion of a rapidly rotating strongly magnetized NS (`magnetar') as an additional energy source (e.g., \citealt{Kangasetal2017, Marguttietal2017, Metzgeretal2017, Nicholletal2017a, Villaretal2017}), although this also promotes the jet feedback scenario \citep{Soker2016Mag,Soker2017Mag2,SokerGilkis2017}. Still, there remains the notable puzzle for jet-driven explosions of the detailed formation process of jets, as well as the question of whether this scenario is limited exclusively to rapidly rotating pre-collapse cores (e.g. \citealt{LeBlanc1970, Khokhlov1999, Wheeleretal2002, Burrowsetal2007, Lazzati2012}) or applicable to \textit{all CCSNe} (the jittering jets explosion scenario, e.g., \citealt{Papish2011, Papish2012b, PapishSoker2014a, PapishSoker2014b, GilkisSoker2014}).

Much less studied is the CITE mechanism, first proposed more than 60 years ago by \cite{Burbidgeetal1957}. It has been recently revived by \cite{KushnirKatz2015}, who suggest that the presence of a mixed helium-oxygen (He-O) shell with certain properties in the pre-collapse core can lead to an energetic CCSN. According to \cite{KushnirKatz2015}, the most energetically favorable composition for their CITE scenario is a helium mass fraction of about $50\%$, and the same mass fraction for oxygen.

As the mass inner to the He-O shell is quite high, a CCSN explosion through the CITE mechanism will generally result in black hole (BH) formation (as suggested for SN 1987A by \citealt{BlumKushnir2016}), with NS formation occurring only with CCSNe of low energy. This introduces one of the shortcomings of the CITE mechanism. For example, \cite{BorkowskiReynolds2017} find a high energy content of $E_\mathrm{SN} \approx 1.9 \times 10^{51} \erg$ for the supernova remnant Kes~73 (G27.4+0.0) as well as a magnetar at its center. Namely, this CCSN that has a NS remnant was quite energetic, contrary to the expectation of the CITE mechanism. Still, the possibility of a CITE ensuing only for some massive stars is not excluded.

Another point in question is the importance of angular momentum. A high core rotation rate will augment the CITE \citep{Kushnir2015a}. The possibility of an accretion disc and jets being formed at the same time in this case has been asserted by \cite{Gilkisetal2016}, and the energy release by jets might outweigh that from nuclear fusion, further complicating matters.

Finally, there is the question of the existence of a He-O shell in the pre-collapse core. The present study addresses this, and to some extent also the issue of angular momentum. It is hard to exaggerate the importance of examining different scenarios for the CCSN phenomenon, and in the present study we set a goal to explore the conditions for the formation of a He-O shell in the pre-collapse core of a massive star and discuss its possible implications. In section \ref{sec:numerical} we describe our numerical set up, and in section \ref{sec:results} we describe the cases where we do find a He-O shell just before core collapse. In section \ref{sec:formation} we discuss the formation or non-formation of a shell with the required He-O composition for an energetic CITE and present in detail several stellar models with mixed shells. In section \ref{sec:summary} we discuss the implications of our findings for any explosion mechanism.

% ==========================================================
\section{NUMERICAL SET UP}
\label{sec:numerical}
% =========================================================

We build a set of rotating stellar models using Modules for Experiments in Stellar Astrophysics (\textsc{mesa} version 8845; \citealt{Paxton2011,Paxton2013,Paxton2015}), with rotation realized in \textsc{mesa} using the ‘shellular approximation’ \citep{Meynet1997}, where the angular velocity $\omega$ is assumed to be constant for isobars. Each model has an initial metallicity of $Z=0.02$ and is evolved from the pre-main sequence stage up to core collapse (`iron' core in-fall velocity of $1000~\mathrm{km}~\mathrm{s}^{-1}$). The models differ in their zero age main sequence (ZAMS) mass and angular velocity. The ZAMS mass range was chosen to be between $M_\zams=13M_\odot$ and $M_\zams=40M_\odot$. We set the ZAMS angular velocity $\omega$ as a fraction of the critical angular velocity and take this fraction to be between $0\%$ and $65\%$. For the initialization of the rotation velocity, the critical angular velocity is defined in \textsc{mesa} as
\begin{equation}
\omega_\mathrm{crit}^2=(1-L/L_\mathrm{Edd})GM/R^3
\label{eq:omegacrit1}
\end{equation}
where $M$ is the total mass, $R$ is the photospheric radius, $L$ is the luminosity, and $L_\mathrm{Edd}$ is the Eddington luminosity of the star (the part of the Eddington luminosity is discussed in detail by \citealt{Maeder2000}).

We computed the critical angular velocity according to equation (\ref{eq:omegacrit1}) for the ZAMS properties of all our models, and found that the ZAMS angular velocity is not exactly the requested fraction of the critical value, apparently due to an averaging over shells performed in the \textsc{mesa} initialization code. This results in a slight difference of the maximal value of $\Omega$, where we define $\Omega \equiv \left(\omega / \omega_\mathrm{crit}\right)_\zams$. For example, requesting $\Omega=0.65$ for the lowest ZAMS mass we take, $13 M_\odot$, yields actually a rotation of $\Omega=0.599$, and for the highest ZAMS mass we take, $40 M_\odot$, we get for the same setting $\Omega=0.506$. Due to this finding, in addition to the intricate role of the Eddington factor \citep{Maeder2000}, we simply present our models in terms of the ZAMS equatorial rotation velocity in units of $\mathrm{km}~\mathrm{s}^{-1}$.

Convection is treated according to the Mixing-Length Theory with $\alpha_\mathrm{MLT}=1.5$. Semiconvective mixing \citep{Langer1983, Langer1991} is employed with $\alpha_{sc}=0.1$. Exponential convective overshooting is applied as in \cite{Herwig2000}, with $f=0.016$ (the fraction of the pressure scale height for the decay scale).

Rotationally-induced mixing is treated as a diffusive process, with diffusion coefficient contributions from dynamical shear instability, Solberg-H{\o}iland instability, secular shear instability, Eddington-Sweet circulation, Goldreich-Schubert-Fricke instability, and the Spruit-Tayler (ST) dynamo (e.g., \citealt{Heger2000,Heger2005}). Rotationally-induced mixing transports angular momentum as well as chemical elements. The ST dynamo \citep{Spruit2002} is less efficient in chemical mixing, and its main effect is reducing the difference in the rotation rates of the core and envelope. Though we did not inhibit the chemical mixing due to the ST dynamo, it is not the single predominant mixing process (Eddington-Sweet circulation and convection are chiefly effective). We also note that according to \cite{Hirschi2004} treatment of the dynamical shear instability is sufficient in order to take into account the Solberg-H{\o}iland instability due to similar timescales, although we did not take special care in that matter. 

Mass-loss during the main sequence phase is treated according to the results of \cite{Vink2001}. During the supergiant phase, mass-loss depends on surface luminosity and temperature according to the fit of \cite{deJager1988}. Some models lose their hydrogen envelope and reach a Wolf-Rayet (WR) phase. At this point mass-loss is according to \cite{WindWR}. The mass-loss rate is rotationally enhanced as in, e.g., \cite{Heger2000}, by a factor of $\left(1-\Omega\right)^{-0.43}$. We did not consider uncertainties in the mass-loss rate (e.g., \citealt{Zilberman2018}).

% ==========================================================
\section{Stellar models with mixed shells of near-equal helium and oxygen mass fractions}
\label{sec:results}
% ==========================================================

We evolve 792 stellar models up to the point of core collapse (or very close to it), with 66 values of the initial rotation velocity for each of the 12 initial masses considered. Fig. \ref{fig:MHeOToV} shows the mass of the He-O shell in all the simulated stellar models. We define a He-O shell as a layer where the helium to oxygen mass ratio is between 0.5 and 2, and the sum of their mass fractions is greater than half. This follows the finding of \cite{Kushnir2015a} that a helium to oxygen mass ratio of 1:1 is energetically optimal, with some allowance for deviation from this ratio, as well as the presence of some elements other than helium and oxygen. To ascertain the outcome of core collapse in the simulated models a more detailed study is required, including hydrodynamic computations with a reaction network. In the present study we focus only on the formation of a mixed He-O shell and the properties of the stellar models for which such a shell is found.
\ifmnras
\begin{figure*}
\else
\begin{figure*}[ht!]
\fi
\centering
\includegraphics[trim= 1.5cm 0.8cm 1cm 1cm,clip=true,width=\textwidth]{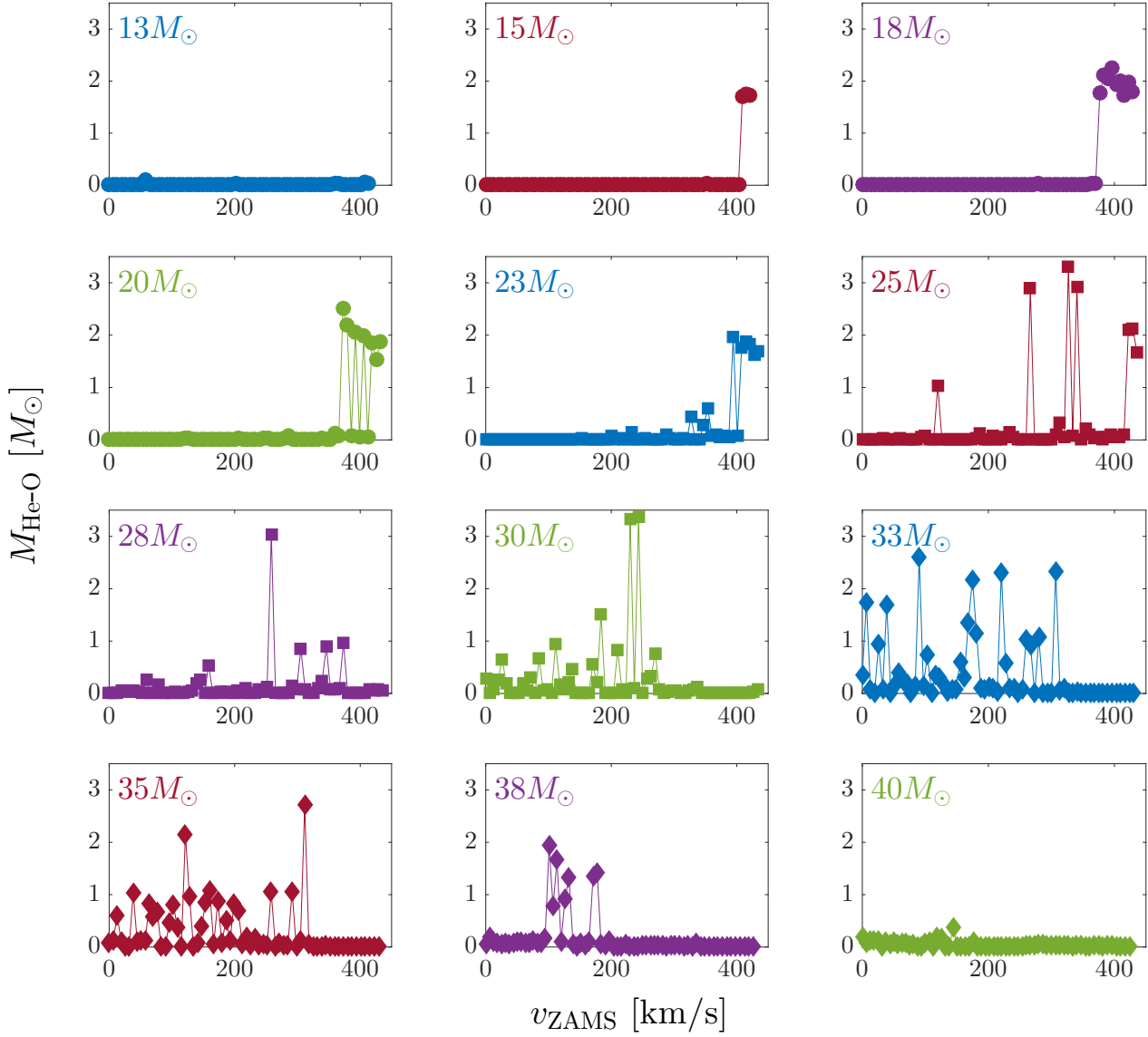}
\caption{The pre-collapse He-O shell mass for various ZAMS masses (indicated for each panel), as function of the ZAMS rotation velocity.}
\label{fig:MHeOToV}
\end{figure*}

The occurrence of a He-O shell is correlated with rotation, and its formation is predominantly due to rotational mixing and convection. It seems that the ST dynamo and Eddington-Sweet circulation are the main rotationally-induced mixing processes. As mentioned in section \ref{sec:numerical}, the ST dynamo might be less effective in mixing chemical elements than in our stellar evolution models \citep{Spruit2002}, but as it is not the sole mixing process, our results might only be a slight overestimate (i.e., the occurrence of a He-O shell might be somewhat less likely).

As one can see from Fig. \ref{fig:MHeOToV}, for $15 M_\odot \le M_\zams \le 23 M_\odot$ only the models with highest initial rotation velocities exhibit a He-O shell. For $30 M_\odot \le M_\zams \le 35 M_\odot$, the manifestation of a He-O shell is more commonplace than for higher or lower masses. We note also that in preliminary simulations with a smaller number of initial velocity values, no He-O shell was found for any model with $M_\zams > 40 M_\odot$.

Fig. \ref{fig:MHeOtoRFinal} plots the He-O shell mass for all models against the stellar photosphere radius. It can be seen that a smaller number of models with a significant He-O shell ($M_\mathrm{He \mbox{-} O} \ga 0.1 M_\odot$) are found with extended envelopes (i.e., supergiants), compared to a larger number of models with a significant He-O shell for which most or all of the hydrogen envelope was lost due to winds.
\ifmnras
\begin{figure}
\else
\begin{figure}[ht!]
\fi
\centering
\includegraphics[trim= 2cm 0.2cm 4cm 1cm,clip=true,width=0.5\textwidth]{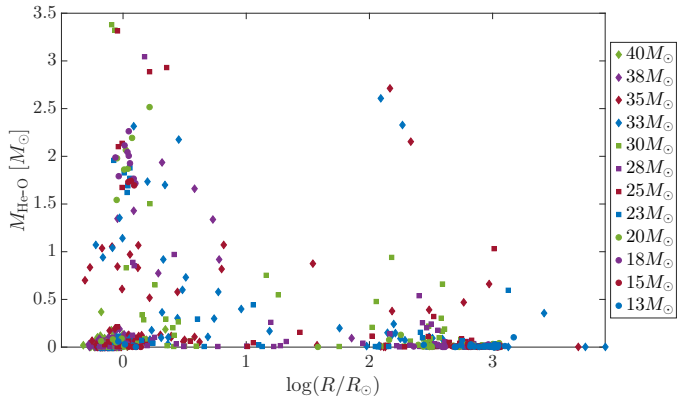}
\caption{The pre-collapse He-O shell mass for different ZAMS rotation velocities and various ZAMS masses as function of the pre-collapse stellar photosphere radius.}
\label{fig:MHeOtoRFinal}
\end{figure}

Fig. \ref{fig:MStratToV} shows that the inner mass coordinate of the He-O shell generally increases monotonically with the ZAMS mass. This is in accordance with the sequence of events leading to the He-O shell formation. First, a separate shell of mostly oxygen is formed beneath a helium shell. The mass coordinates of the helium and oxygen shells naturally correlate with the ZAMS mass. The upper oxygen shell then mixes with the lower part of the helium shell. We attribute the absence of a significant intermediate carbon shell to rotational mixing.
\ifmnras
\begin{figure}
\else
\begin{figure}[ht!]
\fi
\centering
\includegraphics[trim= 2cm 0.2cm 4cm 1cm,clip=true,width=0.5\textwidth]{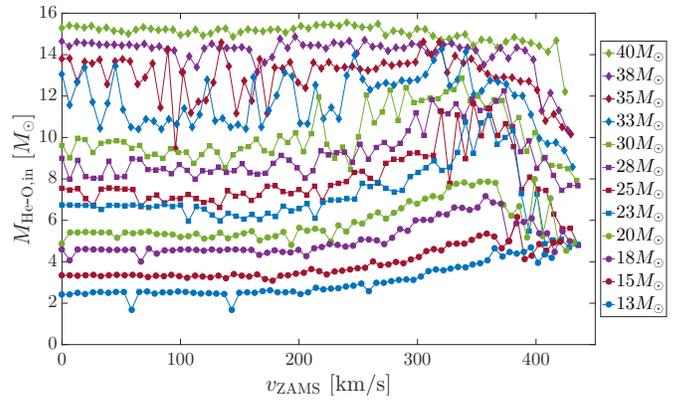}
\caption{The pre-collapse He-O shell inner mass coordinate for various ZAMS masses as function of ZAMS rotation velocity.}
\label{fig:MStratToV}
\end{figure}

Fig. \ref{fig:1523OverTime} shows the development of the He-O shell for $15 M_\odot \le M_\zams \le 23 M_\odot$. For this mass range, the formation of the He-O shell proceeds in a similar manner for all models in which it is found, at about $t_\mathrm{collapse}-t\simeq 100\mathrm{yr}$.
\ifmnras
\begin{figure*}
\else
\begin{figure*}[ht!]
\fi
\centering
\includegraphics[trim= 2cm 0cm 4cm 1cm,clip=true,width=\textwidth]{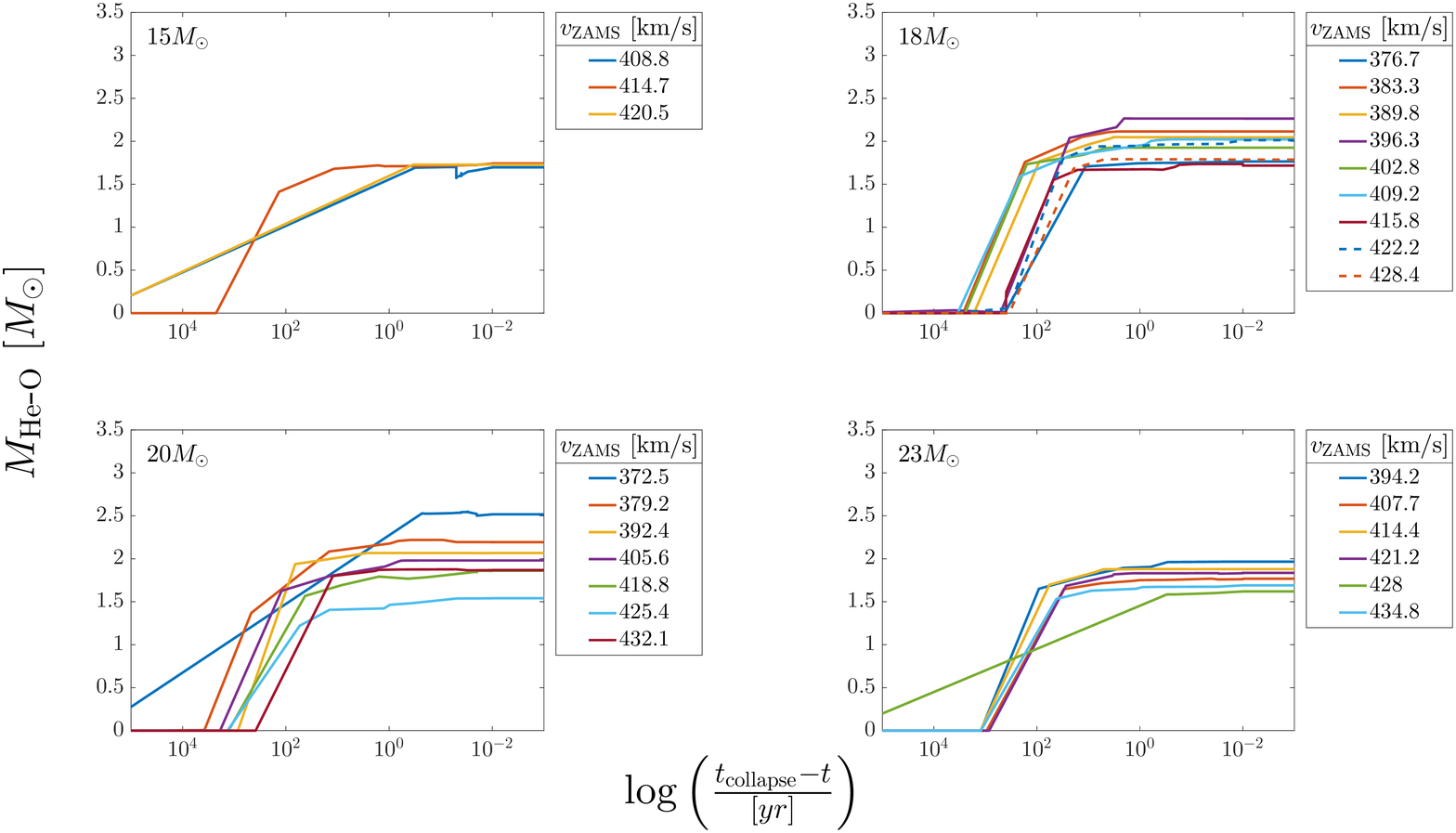}
\caption{The He-O shell mass as a function of time before collapse (i.e., collapse occurs at $t_\mathrm{collapse}-t=0$). This figure presents only models where the pre-collapse He-O shell mass is larger than $0.5 M_{\odot}$.}
\label{fig:1523OverTime}
\end{figure*}

Fig. \ref{fig:jrot} shows the specific angular momentum in the He-O shell for all models with $M_\mathrm{He \mbox{-} O}>0.1 M_\odot$. High specific angular momentum can help augment the explosion energy in the CITE scenario, as the collapsing material has to lose angular momentum, and there is more time for thermonuclear fusion to release enough energy to expel a part of the mixed shell \citep{Kushnir2015a}. To have a notable impact on the collapse dynamics, the specific angular momentum should be comparable to that of the last stable orbit around a BH of a mass equivalent to the material inner to the gas in question. As Fig. \ref{fig:jrot} shows, in most cases the specific angular momentum is only a few percent of that of the innermost stable circular orbit around a Schwarzschild BH with a mass which equals the mass coordinate considered. Still, multidimensional hydrodynamic simulations are necessary to determine the outcome of collapse. If the specific angular momentum is not very high, then a one-dimensional approximation may be appropriate. While \cite{Kushnir2015a} intended to explain all CCSNe through the CITE mechanism, the possibility of sub-luminous CCSNe as an outcome of this scenario should be noted.
\ifmnras
\begin{figure}
\else
\begin{figure}[ht!]
\fi
\centering
\includegraphics[trim= 2cm 0.2cm 4cm 1cm,clip=true,width=0.5\textwidth]{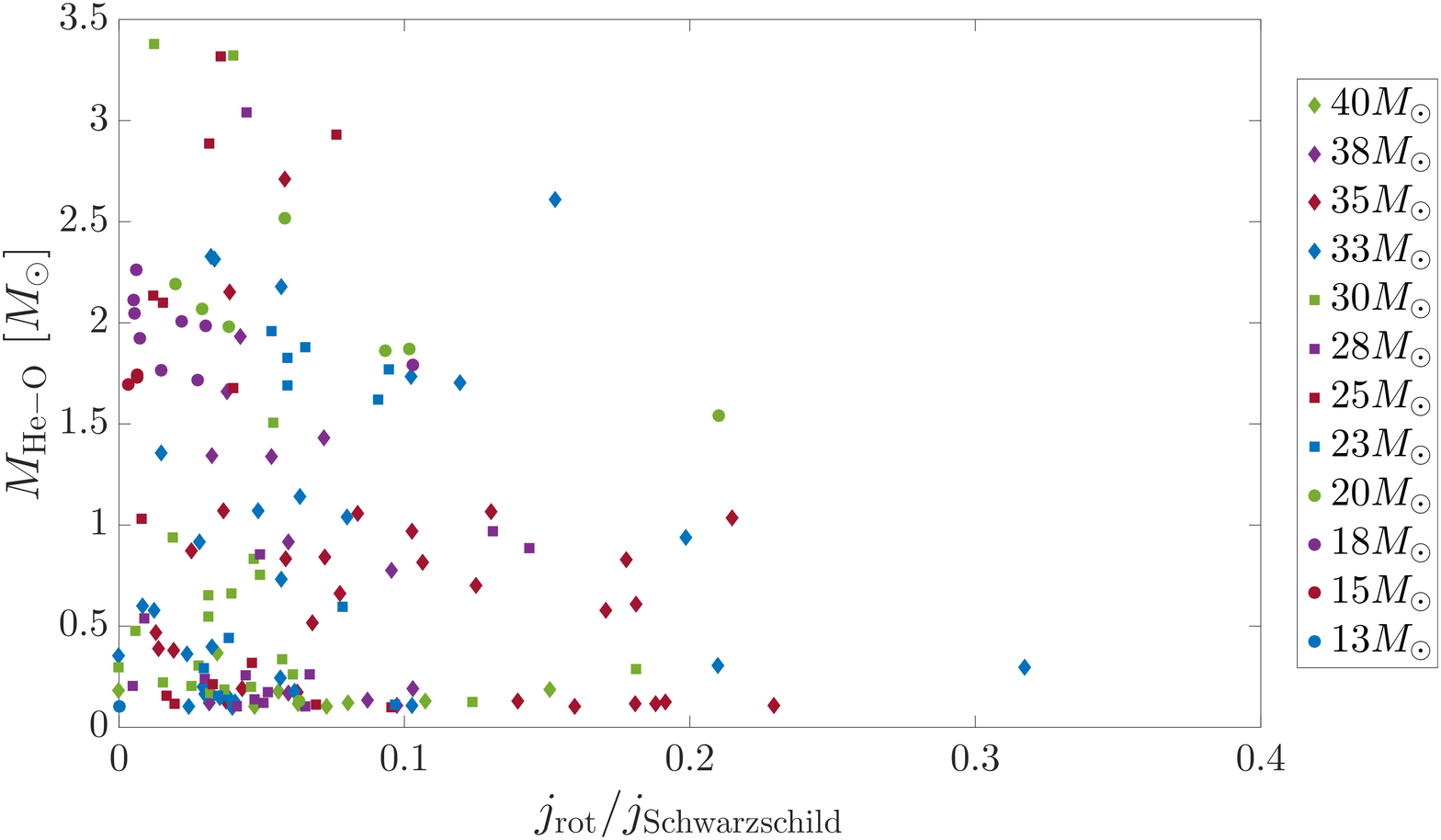}
\caption{The pre-collapse He-O shell mass for different ZAMS rotation velocities and various ZAMS masses as a function of the ratio between $j_\mathrm{rot}$, the specific angular momentum at the middle of the He-O shell and $j_\mathrm{Schwarzschild}$, the specific angular momentum around a Schwarzschild BH with a mass equal to the midpoint mass coordinate in the He-O shell. The graph shows only models in which the pre-collapse mass of the He-O shell is above $0.1 M_\odot$.}
\label{fig:jrot}
\end{figure}

Fig. \ref{fig:MHeOToEnavalopeEnergy} shows the binding energy of the gas above the He-O shell for all models with $M_\mathrm{He \mbox{-} O}>0.1 M_\odot$. This is the  energy needed to lift material from the top of the He-O shell, which we calculate by integration from the surface, taking into account the internal energy as well as the gravitational potential. The majority of the models display a rather low binding energy of the material above the He-O shell, $E_\mathrm{bind} \la 10^{49} \erg$. This is in accordance with low envelope masses, where we define the envelope mass $M_\mathrm{env}$ as the mass above the He-O shell, and we find the following correlation for the models with $M_\mathrm{He \mbox{-} O}>0.1 M_\odot$,
\begin{equation}
\log \left(\frac{M_\mathrm{env}}{M_\odot}\right) \simeq 0.49 + 0.88 \log \left(\frac{E_\mathrm{bind}}{10^{50} \erg} \right).
\label{eq:logMlogE}
\end{equation}
A low-energy CITE might be able to eject the loosely bound envelope. However, not only the binding energy of the gas above the He-O shell should be considered, but also the binding energy of some of the upper part of the mixed shell, which needs to be lifted. The upper half (in terms of mass) of the He-O shell roughly has $E_\mathrm{bind} \ga 10^{50} \erg$.
\ifmnras
\begin{figure}
\else
\begin{figure}[ht!]
\fi
\centering
\includegraphics[trim= 2cm 0.2cm 4cm 1cm,clip=true,width=0.5\textwidth]{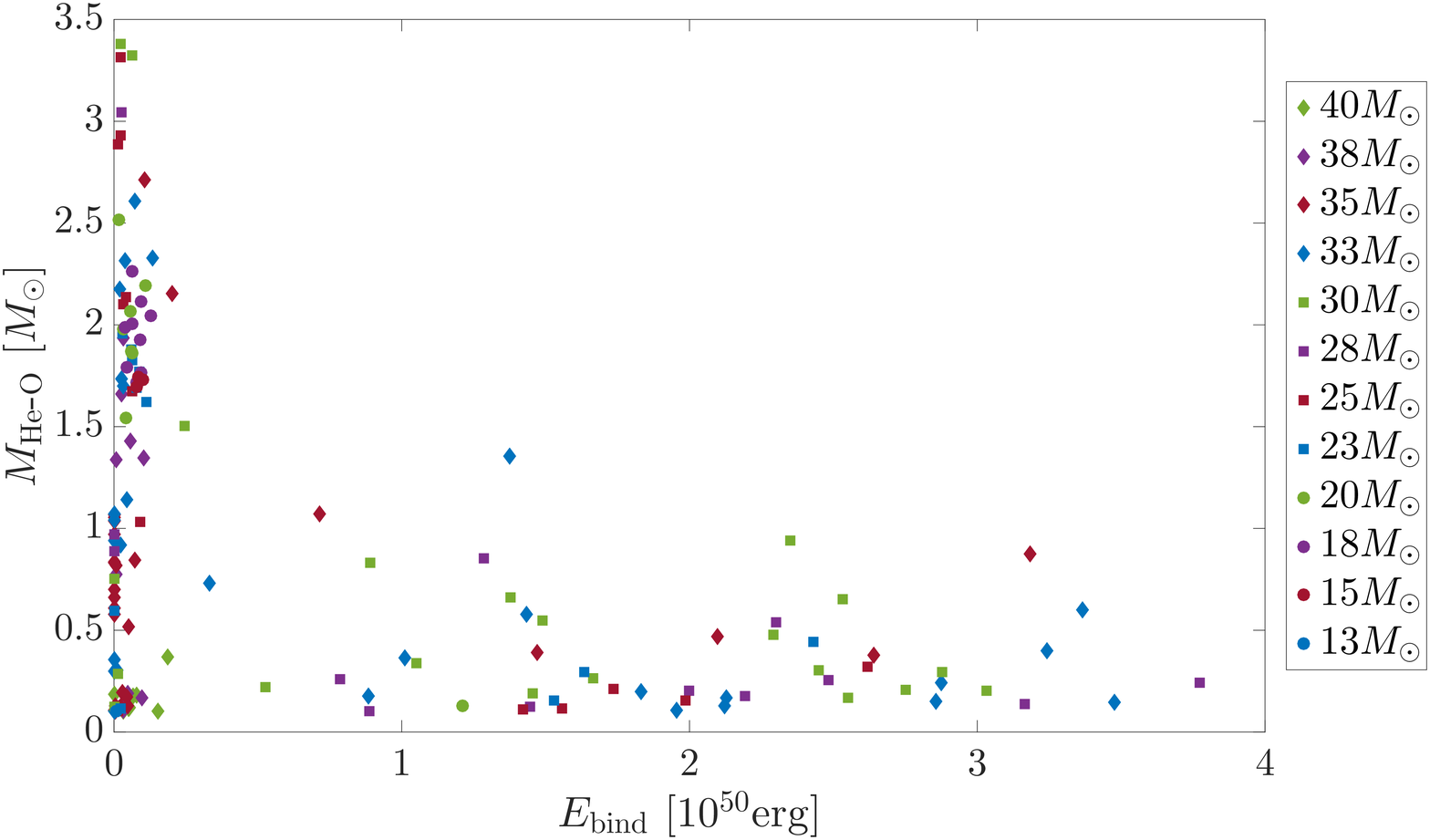}
\caption{The pre-collapse He-O shell mass for different ZAMS rotation velocities and various ZAMS masses as function of the pre-collapse envelope binding energy, where we define the envelope as the gas above the He-O shell (not including it).}
\label{fig:MHeOToEnavalopeEnergy}
\end{figure}

In this section we have focused on the formation of He-O shells where the ratio of mass fractions of helium and oxygen is close to unity. This was motivated by the requirement for such a shell to have a significant influence on the supernova energy if it is ignited \citep{Kushnir2015a}. In the next section we discuss in more detail the formation of mixed shells with various compositions, which do not necessarily have near-equal mass fractions of helium and oxygen, to make a clearer picture of the formation of mixed shells.

% ==========================================================
\section{Mixed shell composition and formation}
\label{sec:formation}
% ==========================================================

In this section we present in detail several stellar models to show the formation and composition of mixed shells. In Fig. \ref{fig:HighestHeOM} and Fig. \ref{fig:RGBHeO} we present the internal composition structure of two stellar models in detail. We chose to present the model with the highest He-O shell mass, and one model which has a significant He-O shell and is also a red supergiant. Fig. \ref{fig:HighestHeOM} presents the model with the highest He-O shell mass according to our definition. The He-O shell resides between the mass coordinates $M_\mathrm{He \mbox{-} O,in}=8.46 M_\odot$ and $M_\mathrm{He \mbox{-} O,out}=11.84 M_\odot$, the helium to oxygen ratio is approximately 1:2, and the sum of their mass fractions is $f_\mathrm{He \mbox{-} O} \approx 0.73$. Some carbon and neon are present in the shell as well. %The detailed reactions for such mixed compositions will be studied in future work.
\ifmnras
\begin{figure}
\else
\begin{figure}[ht!]
\fi
\centering
\includegraphics[trim= 1cm 0.2cm 4cm 1cm,clip=true,width=0.5\textwidth]{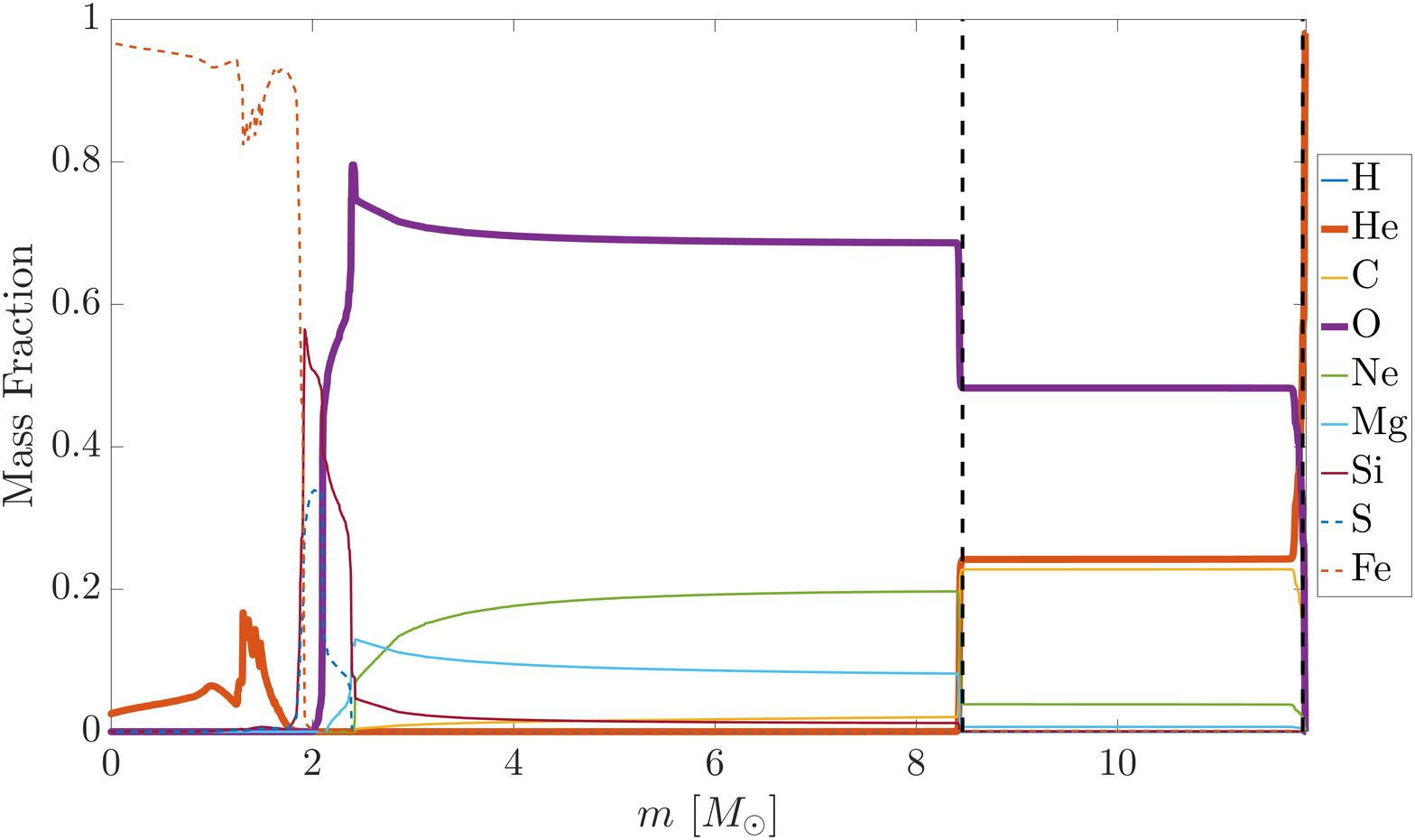}
\caption{Composition as function of mass coordinate at the pre-collapse stage, for the model with the highest pre-collapse He-O shell mass, of $M_\mathrm{He \mbox{-} O}=3.38M_\odot$. The initial parameters of the model are $M_\zams = 30 M_\odot$ and $v_\zams = 243.45 ~\mathrm{km} ~\mathrm{s}^{-1}$. The He-O shell according to our definition is marked by two horizontal dashed black lines. The outer edge of the He-O shell is very close to the photosphere radius, $R=0.81 R_\odot$. %$R=10^{-0.0914}R_\odot$ $R=0.8102R_\odot$
}
\label{fig:HighestHeOM}
\end{figure}
\ifmnras
\begin{figure}
\else
\begin{figure}[ht!]
\fi
\centering
\includegraphics[trim= 2cm 0.2cm 4cm 1cm,clip=true,width=0.5\textwidth]{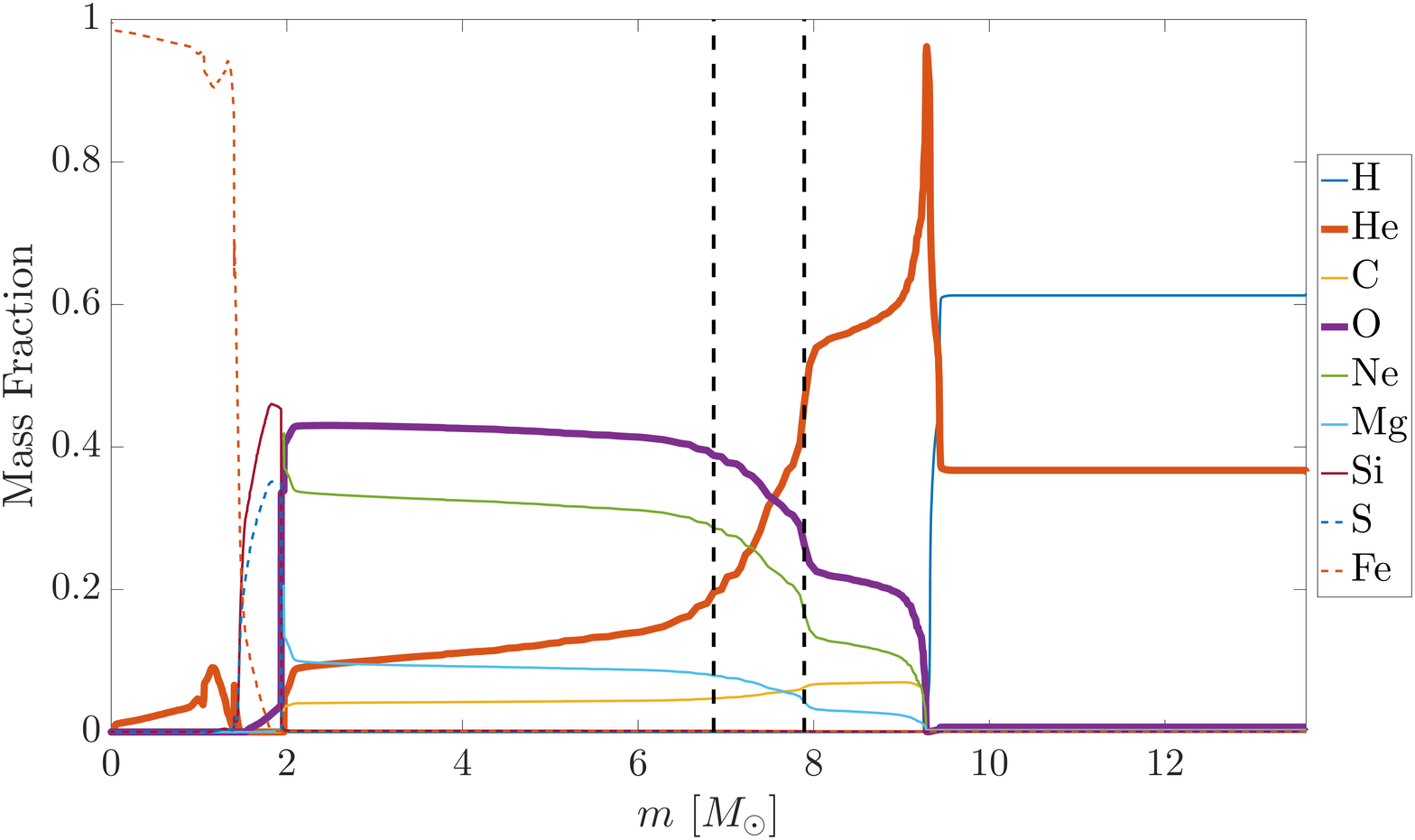}
\caption{Composition as function of mass coordinate at the pre-collapse stage, for a model with a pre-collapse photosphere radius of $R=1030 R_\odot$ %$R= 10^{3.01} R_\odot = 1030.2 R_\odot$ 
which is suitable for a red supergiant. The initial parameters of the model are $M_\zams = 25 M_\odot$ and $v_\zams = 120.3 ~\mathrm{km} ~\mathrm{s}^{-1}$.  The He-O shell according to our definition is marked by two horizontal dashed black lines.}
\label{fig:RGBHeO}
\end{figure}

Fig. \ref{fig:RGBHeO} presents a model which is a red supergiant at the pre-collapse stage, and has a He-O shell mass of $M_\mathrm{He \mbox{-} O}=1.03M_\odot$ according to our definition. The model has a mixed He-O shell according to our definition in section \ref{sec:results}, though with a composition different than in the model presented in Fig. \ref{fig:HighestHeOM}. The helium and oxygen comprise a fraction of $f_\mathrm{He \mbox{-} O} \approx 0.68$ of the shell in terms of mass, with a presence of some neon, carbon and magnesium as well. The binding energy of the hydrogen envelope above the He-O shell is $E_\mathrm{bind} \approx 10^{49} \erg$. Above the region we define as a He-O shell, there is a mixture where the helium to oxygen ratio is roughly 3:1. This shows a minor limitation of our definition, as such a mixture might also yield an explosive output. %While our study provides a general estimate for the formation of He-O shells in CCSN progenitors, more detailed calculations are needed to ascertain the outcome of collapse, including an appropriate reaction network.

Fig. \ref{fig:Kippenhan1} shows the evolution over time of the internal composition structure of two models with $M_\zams = 20 M_\odot$, with one model having an initial rotation velocity of $v_\zams=379.17 ~\mathrm{km} ~\mathrm{s}^{-1}$ and the other $v_\zams=385.78 ~\mathrm{km} ~\mathrm{s}^{-1}$. In the former a significant mixed shell (according to our definition in section \ref{sec:results}) forms a few centuries before core collapse, while for the latter a mixed shell according to the same definition does not form.
\ifmnras
\begin{figure}
\else
\begin{figure}[ht!]
\fi
\centering
\includegraphics[trim= 2cm 0.2cm 4cm 0.7cm,clip=true,width=0.5\textwidth]{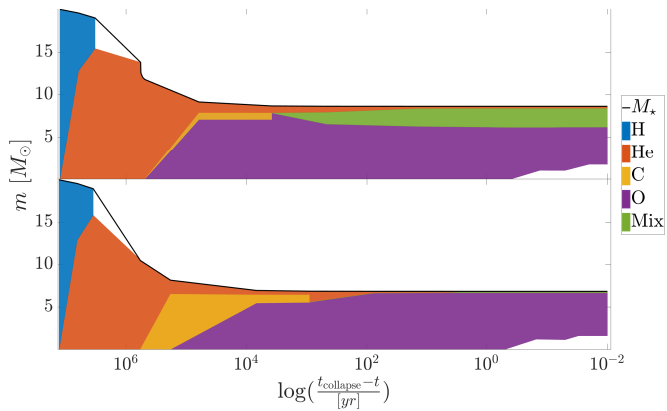}
\caption{Kippenhahn diagram for a model where a pre-collapse He-O shell forms (top) according to our requirement of helium to oxygen mass ratio in the range $0.5$-$2$, and a model where a shell with that mass ratio does not form (bottom). Both models have a ZAMS mass of $M_\zams = 20 M_\odot$, but the initial rotation velocities for the top and bottom panels are $v_\zams=379.17 ~\mathrm{km} ~\mathrm{s}^{-1}$ and $v_\zams=385.78 ~\mathrm{km} ~\mathrm{s}^{-1}$, respectively.}
\label{fig:Kippenhan1}
\end{figure}

We now look into the reason for a mixed shell forming in one model and not in another, even though their initial conditions are very similar. We redefine the helium to oxygen ratio required for the presence of a mixed shell as $r_\mathrm{He \mbox{-} O}$, so that a region of the star is part of a mixed shell if
\begin{equation}
r_\mathrm{He \mbox{-} O} \ge f_\mathrm{He}/f_\mathrm{O} \ge 1/ r_\mathrm{He \mbox{-} O},
\label{eq:ratio}
\end{equation}
where $f_\mathrm{He}$ and $f_\mathrm{O}$ are the helium and oxygen mass ratios, respectively, and we keep the requirement that $f_\mathrm{He}+f_\mathrm{O} \ge 0.5$. In section \ref{sec:results} we used $r_\mathrm{He \mbox{-} O}=2$ throughout, as well as in Figs. \ref{fig:HighestHeOM}-\ref{fig:Kippenhan1}. In Fig. \ref{fig:Kippenhan2} we show for comparison the evolution over time of the internal composition structure for $r_\mathrm{He \mbox{-} O}=5$, i.e., for the same internal composition structure as in Fig. \ref{fig:Kippenhan1} but with a mixed shell defined also for a larger asymmetrical ratio between the helium and oxygen mass ratios. We see then that a mixed shell forms in both models. For the $v_\zams=385.78 ~\mathrm{km} ~\mathrm{s}^{-1}$ model, though, the mass ratio between helium and oxygen is farther from unity than for the $v_\zams=379.17 ~\mathrm{km} ~\mathrm{s}^{-1}$ model.
\ifmnras
\begin{figure}
\else
\begin{figure}[ht!]
\fi
\centering
\includegraphics[trim= 2cm 0.2cm 4cm 0.7cm,clip=true,width=0.5\textwidth]{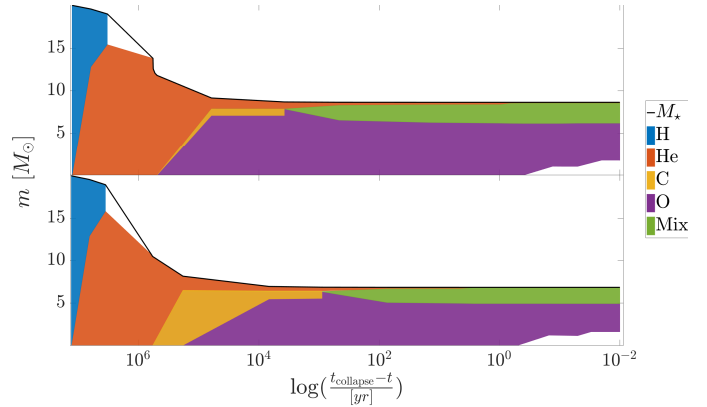}
\caption{Like Fig. \ref{fig:Kippenhan1}, but now the mixed He-O layer is defined for a helium to oxygen mass ratio in the range $0.2$-$5$ ($r_\mathrm{He \mbox{-} O}=5$ in equation \ref{eq:ratio}). }
\label{fig:Kippenhan2}
\end{figure}

In Fig. \ref{fig:Comp1} we present in detail the internal structure of the two models whose evolution is shown in Figs. \ref{fig:Kippenhan1} and \ref{fig:Kippenhan2}, at a similar evolutionary phase. Both models are at the stage of shell carbon burning, and their structure is that of a carbon-oxygen core and a helium-rich envelope, with a mixed shell between them. The models differ in the composition of the mixed shell: The mixed shell in the slightly slower model (top two panels) has a composition of $41.8\%$ carbon, $33.1\%$ oxygen and $22.7\%$ helium, while the faster model (bottom two panels) has in its mixed shell $42.7\%$ carbon, $38.2\%$ oxygen and $16.7\%$ helium. In both cases, heavier elements amount to the remaining $2.4\%$ of the composition, most of which ($1.8\%$) is neon.
\ifmnras
\begin{figure*}
\else
\begin{figure*}[ht!]
\fi
\centering
\includegraphics[trim= 0.2cm 0.5cm 0cm 0cm,clip=true,width=\textwidth]{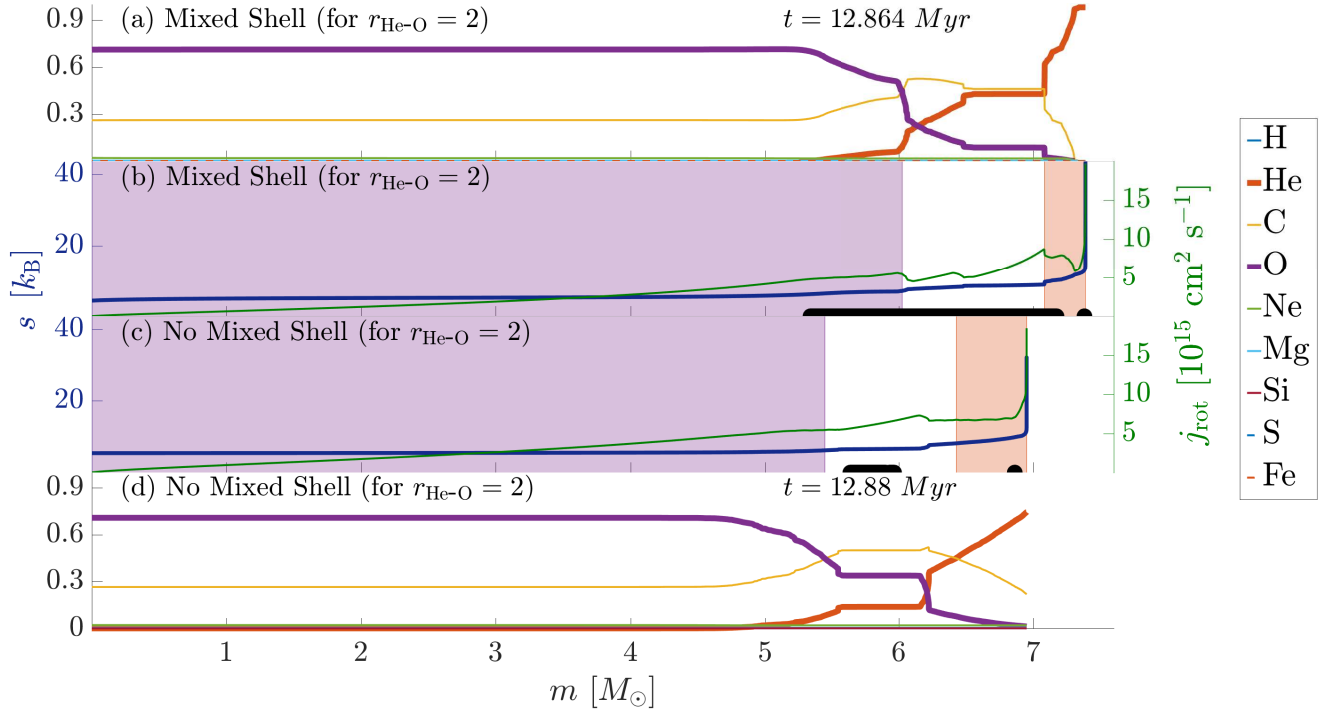}
\caption{Comparison between a model where a pre-collapse He-O shell  (for $r_\mathrm{He \mbox{-} O} = 2$) forms and a model where such a shell does not form. Both models have a ZAMS mass of $M_\zams = 20 M_\odot$ and are presented at the same evolutionary stage where carbon burns in a shell above a core composed mostly of oxygen. The top two panels present the model with an initial rotation velocity of $v_\zams=379.17 ~\mathrm{km} ~\mathrm{s}^{-1}$ where a pre-collapse He-O shell forms. The two bottom panels belong to the model with an initial rotation velocity of $v_\zams = 385.78 ~\mathrm{km} ~\mathrm{s}^{-1}$ where a pre-collapse He-O shell does not form (according to $r_\mathrm{He \mbox{-} O} = 2$). \textit{Panels (a) + (d):} The composition as a function of the mass coordinate. \textit{Panels (b) + (c):} The blue line (left axis) is the specific entropy as a function of mass coordinate. The green line (right axis) is the specific angular momentum. The orange background defines the helium shell and the purple the oxygen shell. The black marks at the bottom illustrate the area where convection occurs.}
\label{fig:Comp1}
\end{figure*}

Another important feature shown in Fig. \ref{fig:Comp1} is that the mixed shell is partly or completely convective. This results in the uniform composition throughout most of the mixed shell for the $v_\zams = 385.78 ~\mathrm{km} ~\mathrm{s}^{-1}$ model. For the $v_\zams=379.17 ~\mathrm{km} ~\mathrm{s}^{-1}$ model, the composition will become uniform throughout the mixed shell further along its evolution due to convective mixing.

In Fig. \ref{fig:MassRatio1} we further show that our analysis of whether a mixed shell forms or not depends on our criterion of a mixed shell in terms of the mass ratio between helium and oxygen. When the ratio is allowed to be in a larger range than that used in section \ref{sec:results} for us to claim a mixed shell exists, then more models have a mixed shell, and the ``noisy'' behavior seen in some cases in Fig. \ref{fig:MHeOToV} disappears\footnote{The ``noise'' does not disappear completely, though this is not unexpected, as the details of the last stages of massive stellar evolution depend sensitively on the initial conditions. For example, \cite{Sukhboldetal2018} show that the pre-collapse core structure varies irregularly as a function of the initial mass. Slight variations in the initial rotation velocity can have similar effects, as rotational mixing brings more fuel to the core, and a stellar model might then behave similar to one with a slightly higher initial mass.}. The earlier requirement for a low value of $r_\mathrm{He \mbox{-} O}$ arises from the advantage of a mass ratio close to unity in terms of the CITE mechanism \citep{Kushnir2015a}.
\ifmnras
\begin{figure}
\else
\begin{figure}[ht!]
\fi
\centering
\includegraphics[trim= 2.1cm 1cm 0.8cm 0.3cm,clip=true,width=0.5\textwidth]{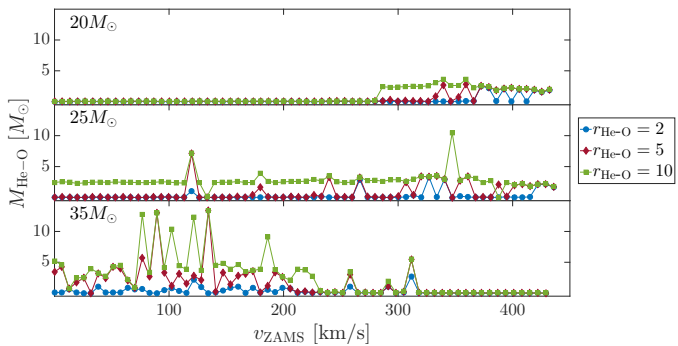}
\caption{ The pre-collapse He-O shell mass for models with $M_\zams = 20 M_\odot$ (top), $M_\zams = 25 M_\odot$ (middle) and $M_\zams = 35 M_\odot$ (bottom) as a function of the ZAMS rotation velocity and for three criteria for the definition of a mixed shell. The blue circle, red diamond and green square markers are for a helium to oxygen mass ratio in the range $0.5$-$2$ ($r_\mathrm{He \mbox{-} O}=2$ in equation \ref{eq:ratio}), $0.2$-$5$ ($r_\mathrm{He \mbox{-} O}=5$) and $0.1$-$10$ ($r_\mathrm{He \mbox{-} O}=10$), respectively.}
\label{fig:MassRatio1}
\end{figure}

From the analysis presented above, and further such analysis for numerous other models (not presented for the sake of brevity), a coherent picture of the mixed-shell formation emerges. A mixed shell forms in many cases, but its composition varies. In many cases the mixed shell has a large fraction of carbon, and we do not consider it as having a promising composition within the framework of the CITE mechanism for CCSNe\footnote{ \cite{KushnirKatz2015} briefly mention a helium-carbon mixture as a viable option also leading to a CITE, but they claim the mixture of helium and oxygen to be more favorable. The explosive outcome from a mixture of helium, carbon, and oxygen should be studied in the future. Still, the specific angular momentum in the mixed shell is probably too low to allow the CITE to operate.}. It appears there is a threshold initial rotation velocity for the formation of a mixed shell, which becomes lower with increasing mass. Starting at $M_\zams \approx 30 M_\odot$, high values of the initial rotation velocities lead to excessive mass-loss, so that almost no helium is left in the star at all. A mixed helium-oxygen shell then does not form.

% ==========================================================
\section{DISCUSSION AND SUMMARY}
\label{sec:summary}
% ==========================================================

This study is motivated by the disagreement on the explosion mechanism of massive stars. One mechanism involves nuclear fusion of the collapsing core material that releases enough energy to explode the star \citep{Burbidgeetal1957}, in particular the nuclear fusion of helium and oxygen, in a mixed helium-oxygen (He-O) shell \citep{Kushnir2015a, KushnirKatz2015}. 

Using the numerical stellar evolution code \textsc{mesa} we have simulated the evolution of 792 massive stellar models to the point of core collapse, and searched for the presence of a mixed He-O shell. Based on the requirements of this collapse-induced thermonuclear explosion (CITE) mechanism, we consider a star to have such a He-O shell if the helium to oxygen mass ratio in the shell is between 0.5 and 2, and if the combined fractions of their masses in that shell is greater than half. We summarize our main results and their implications as follows.  

(1) The formation of a He-O shell according to the definition of section \ref{sec:results} is rare (Fig. \ref{fig:MHeOToV}), and in most cases the mass of the He-O shell is lower than what is required by the CITE mechanism for an energetic CCSN (Fig. \ref{fig:MHeOtoRFinal}). Out of 792 stellar evolutionary models that we simulated, only in 160 (86) models we have found a He-O shell mass of $> 0.1 M_\odot$ ($> 0.5 M_\odot$). For models in the lower part of the mass range, that supply most of the CCSNe, the incidence is lower even. We have found that only $1.52 \%$ of the stars with ZAMS masses of $13 M_\odot$ and $4.55 \%$ of the stars with ZAMS masses of $15 M_\odot$ have He-O shells (for $r_\mathrm{He \mbox{-} O}=2$ in Eq. \ref{eq:ratio} and with a He-O shell mass of $> 0.1 M_\odot$) before explosion. Since it is expected that most massive stars explode as supernovae, we conclude that the nuclear burning of the He-O shell (the CITE mechanism) cannot be the main explosion mechanism of CCSNe. 
 
(2) Another problem the CITE mechanism encounters is that it requires the pre-collapse core to have a high specific angular momentum \citep{Kushnir2015a}. We found that the pre-collapse cores have a lower  angular velocity than what the CITE mechanism requires for an energetic explosion (Fig. \ref{fig:jrot}).

(4) From Fig. \ref{fig:MHeOToEnavalopeEnergy} we learn that very massive He-O shells, namely, $M_{\rm He \mbox{-} O} \ga 1 M_\odot$ are formed in pre-explosion models with very low envelope binding energies. The explanation is that such massive shells are formed in models that have lost most of their envelope (Eq. \ref{eq:logMlogE}). This does not necessarily imply that the CITE mechanism, if takes place, in these cases leads to CCSNe with very high ejecta velocities. The reason is that a successful energetic CITE requires rapid rotation and many of the models with massive He-O shells do not rotate rapidly (Fig. \ref{fig:jrot}). Another possibility is that such stars with massive He-O shells will lead to peculiar sub-luminous CCSNe of type Ib or Ic.

(5) From Fig. \ref{fig:MHeOToEnavalopeEnergy} we also learn that in many models the He-O shell mass is relatively low, $M_{\rm He \mbox{-} O} \la 0.5 M_\odot$ and the envelope binding energy is $\ga 10^{50} \erg$. To prevent a fine tuning we require that the explosion energy will be about twice or more of the binding energy. Since in the CITE mechanism part of He-O shell falls on to the center \citep{Kushnir2015a, KushnirKatz2015}, it is questionable whether in these cases the CITE mechanism can work at all. If it works, the kinetic energy of the ejecta will be very low, i.e., $\la 10^{50} \erg$.

(6) \cite{WoosleyWeaver1995}, \cite{Hashimoto1995}, and \cite{Nakamuraetal2001}, among others, calculated the nucleosynthesis resulting from the propagation of the exploding shock through the cores of massive stars. Their results show that only the inner part of the oxygen shell undergoes substantial nuclear reactions. \cite{Nakamuraetal2001} studied the role of he explosion energy in the range of $10^{51} \erg$ to $10^{53} \erg$, and found that for more energetic explosions larger fractions of the oxygen shell suffer nuclear reactions. This behavior implies that in the majority of cases not much nucleosynthesis takes place in the He-O shell, because the He-O shell mainly resides in the outer part of the oxygen shell (Figs.  \ref{fig:MStratToV} and \ref{fig:HighestHeOM}). In rare cases where the He-O shell inner boundary is very close to the inner boundary of the oxygen shell (although in our models the minimal separation is $M_\mathrm{He \mbox{-} O,in} - M_\mathrm{O,in} \ga 2 M_\odot$) and/or in rare cases of very energetic explosion of $E_{\rm exp} \ga 10^{52} \erg$ we expect that the propagating shock will set some nuclear reactions in the He-O shell. More likely are cases where propagating jets penetrate the He-O shell and in small regions induce nucleosynthesis by shocking the He-O mixture. Future studies should address this possibility of jet-induced nucleosynthesis in a He-O shell.  

(7) The formation of a mixed shell is a general trend, with a rotation velocity threshold which decreases with increasing initial mass. The composition of the mixed shell, though, varies. In many cases the mixed shell has a high ratio between the mass fraction of helium and the mass fraction of oxygen, or otherwise it has a significant amount of carbon (section \ref{sec:formation}). The explosive (or non-explosive) nature of mixed shells with different compositions should be studied in the future, though in this work we focused on mixed shells composed mostly of helium and oxygen, following the claim of \cite{KushnirKatz2015} that such a mixture is the most favorable for the CITE mechanism.

Overall, the presence of a mixed helium-oxygen shell is rare, and even in cases where it does exist in the pre-collapse core its mass might be too low and/or the rotation too slow to lead to an even weak explosion. We speculate that the main effect of a He-O shell might be in cases where it is shocked by jets and undergoes some nuclear reactions that are less likely to take place without this shell. This should be the topic of a future study. 

\section*{Acknowledgments}

We thank an anonymous referee for helpful suggestions.
We acknowledge support from the Israel Science Foundation and a grant from the Asher Space Research Institute at the Technion. AG thanks the support of the Blavatnik Family Foundation.

\ifmnras
\bibliographystyle{mnras}

\label{lastpage}
\end{document}